\documentclass[twocolumn,aps,pra,showpacs,superscriptaddress]{revtex4}

\usepackage{graphicx}
\usepackage{amsmath,amssymb}
\usepackage{color}

\begin{document}

%%%%%%%%%%%%%%%%%%%%%%%%%%%%%%%%%%%%%%%%%%%%%%%%%%%%%%%%%%%%%%%%%%%

\title{Bell inequalities for the simplest exclusivity graph}

%%%%%%%%%%%%%%%%%%%%%%%%%%%%%%%%%%%%%%%%%%%%%%%%%%%%%%%%%%%%%%%%%%%

%sadiq@fysik.su.se,piotr.badziag@gmail.com,boure@fysik.su.se

\author{Muhammad Sadiq}
 %\email{sadiq@fysik.su.se}
 \affiliation{Department of Physics, Stockholm University, S-10691, Stockholm, Sweden}

\author{Piotr Badzi{\c a}g}
 %\email{pbg01@fysik.su.se}
\affiliation{Department of Physics, Stockholm University, S-10691, Stockholm, Sweden}

\author{Mohamed Bourennane}
%\email{boure@fysik.su.se}
\affiliation{Department of Physics, Stockholm University, S-10691, Stockholm, Sweden}

\author{Ad\'an Cabello}
%\email{adan@us.es}
\affiliation{Departamento de F\'{\i}sica Aplicada II, Universidad de Sevilla, E-41012 Sevilla, Spain}
\affiliation{Department of Physics, Stockholm University, S-10691, Stockholm, Sweden}

%%%%%%%%%%%%%%%%%%%%%%%%%%%%%%%%%%%%%%%%%%%%%%%%%%%%%%%%%%%%%%%%%%%

\date{\today}

%First version: March 18, 2011 (Stockholm).
%Previous version: January 12, 2013 (Concepción).
%This version: January 27, 2013 (Belo Horizonte). After PRA proofs.

%%%%%%%%%%%%%%%%%%%%%%%%%%%%%%%%%%%%%%%%%%%%%%%%%%%%%%%%%%%%%%%%%%%

\begin{abstract}
Which is the simplest logical structure for which there is quantum nonlocality? We show that there are only three bipartite Bell inequalities with quantum violation associated with the simplest graph of relationships of exclusivity with a quantum-classical gap. These are the most elementary logical Bell inequalities. We show that the quantum violation of some well-known Bell inequalities is related to them. We test the three Bell inequalities with pairs of polarization-entangled photons and report violations in good agreement with the quantum predictions. Unlike other experiments testing noncontextuality inequalities with pentagonal exclusivity, the ones reported here are free of the compatibility loophole.
\end{abstract}

%%%%%%%%%%%%%%%%%%%%%%%%%%%%%%%%%%%%%%%%%%%%%%%%%%%%%%%%%%%%%%%%%%%

%Optical tests of quantum theory

\maketitle

%%%%%%%%%%%%%%%%%%%%%%%%%%%%%%%%%%%%%%%%%%%%%%%%%%%%%%%%%%%%%%%%%%%

\section{Introduction}

%%%%%%%%%%%%%%%%%%%%%%%%%%%%%%%%%%%%%%%%%%%%%%%%%%%%%%%%%%%%%%%%%%%

Contextual and nonlocal correlations, defined as those that cannot be reproduced with noncontextual hidden variable (NCHV) \cite{Specker60,Bell66,KS67} and local hidden variable (LHV) theories \cite{Bell64}, are valuable resources for information processing. In an attempt to identify potentially interesting experimental setups, one usually investigates whether the set of correlations in a given experimental scenario has points outside the polytope of noncontextual or local correlations \cite{Fine82,Pitowsky89}. To characterize the boundaries of these polytopes and, consequently, to find the interesting scenarios is, however, computationally hard \cite{Pitowsky89}.

To circumvent some of the difficulties, one can link noncontextual (NC) inequalities to graph theory \cite{CSW10}. In this approach, every NC inequality is expressed in terms of a convex combination $S$ of probabilities of events $a,b,\ldots | x,y, \ldots$, denoting ``compatible tests $x,y,\ldots$ gave results $a,b,\ldots$.'' Each of these events can be represented by a vertex of a graph $G$ whose edges link exclusive events (i.e., those which cannot be simultaneously true). It has been shown \cite{CSW10} that the maximum value of $S$ for LHV and NCHV theories is given by the graph's independence number $\alpha(G)$ \cite{Diestel10}, while the maximum in quantum mechanics (QM) is upper bounded by the graph's Lov\'asz number $\vartheta(G)$ \cite{Lovasz79}. Therefore, NC inequality violated by QM can be associated with a graph $G$ such that $\alpha (G) < \vartheta (G)$. Conversely, for any graph $G$ such that $\alpha (G) < \vartheta (G)$, there is NC inequality for which the upper bound of $S$ for any LHV and NCHV theory is $\alpha(G)$ and the maximum in QM is exactly $\vartheta(G)$. This suggests a fundamental role of the exclusivity structures in the quantum advantage. This approach has been used to identify tasks \cite{NDSC12} and NC inequalities \cite{ADLPBA11} for which QM outperforms NCHV theories.

The graph approach to NC inequalities singles out the pentagon $C_5$ as the graph with the least number of vertices, for which $\alpha(G) < \vartheta(G)$ \cite{CSW10,Diestel10,Lovasz79}. Its exclusivity structure was already discussed by Wright \cite{Wright78} in the 1970s. Moreover, $C_5$ is the graph behind a fundamental NC inequality for a single qutrit identified by Klyachko, Can, Binicioglu, and Shumovski (KCBS) \cite{KCBS08,BBCGL11}, and tested in experiments with sequential measurements \cite{LLSLRWZ11,AABC11}.

Local realism described by Bell inequalities can be regarded as a restricted form of NC, where compatibility of measurements is replaced by a more stringent requirement of spatial separation. The extra restriction may make linking graphs and Bell inequalities less straightforward than linking graphs and NC inequalities. An interesting if not a fundamental question is thus to what extent Bell inequalities can be described by graphs with equal success as NC inequalities.

With this question in mind, in this article we find that, even for Bell inequalities, the pentagon represents the simplest logical structure for which there is a classical-quantum gap. Subsequently, we show that there are three nonequivalent bipartite Bell inequalities represented by the exclusivity structure of $C_5$, and we link these (pentagonal) inequalities to well-known Bell inequalities, Clauser-Horne-Shimony-Holt (CHSH) inequality \cite{CHSH69} and $I_{3322}$ \cite{Froissart81,Sliwa03,CG04}, and to the KCBS inequality. We test the inequalities experimentally. Via the link to the KCBS inequality, our experiments provide experimental proofs of the violation of pentagonal NC inequalities free of the compatibility loophole \cite{GKCLKZGR10}. Moreover, due to recent theoretical results \cite{A12,Cabello12}, our link between pentagonal Bell inequalities and the CHSH inequality provides particularly simple logical evidence for the impossibility of the Popescu-Rohrlich (PR) nonlocal boxes \cite{PR94}.

%%%%%%%%%%%%%%%%%%%%%%%%%%%%%%%%%%%%%%%%%%%%%%%%%%%%%%%%%%%%%%%%%%%

\section{Scenario}

%%%%%%%%%%%%%%%%%%%%%%%%%%%%%%%%%%%%%%%%%%%%%%%%%%%%%%%%%%%%%%%%%%%

Our first concern was the question of whether the pentagonal structure $C_5$ obeying the bipartite limitation produces a quantum-classical gap. To model this scenario, the vertices of our pentagon represent events $ab|xy$ associated with two distant parties: Alice testing $x$ and obtaining $a$, and Bob testing $y$ and obtaining $b$. Due to the separation between Alice and Bob, the edges of the pentagon link $ab|xy$ and $a'b'|x'y'$ if and only if $x=x'$ and $a \neq a'$ or $y=y'$ and $b \neq b'$. With $P(ab|xy)$ denoting the probability of $ab|xy$, one can associate the pentagon with Bell inequalities of the following form:
\begin{equation}
 \label{zero}
 \sum_{ab|xy \in C_5} P(ab|xy) \stackrel{\mbox{\tiny{LHV}}}{\leq} \Omega_{\mathrm{LHV}} \stackrel{\mbox{\tiny{QM}}}{<} \Omega_{\mathrm{QM}},
\end{equation}
where $\stackrel{\mbox{\tiny{LHV}}}{\leq} \Omega_{\mathrm{LHV}}$ indicates that $\Omega_{\mathrm{LHV}}$ is the supremum for every LHV theory, and $\stackrel{\mbox{\tiny{QM}}}{<} \Omega_{\mathrm{QM}}$ indicates that the supremum in QM $\Omega_{\mathrm{QM}}$ is strictly larger.

%%%%%%%%%%%%%%%%%%%%%%%%%%%%%%%%%%%%%%%%%%%%%%%%%%%%%%%%%%%%%%%%%%%
%Fig. 1
%%%%%%%%%%%%%%%%%%%%%%%%%%%%%%%%%%%%%%%%%%%%%%%%%%%%%%%%%%%%%%%%%%%

\begin{figure}[t]
 \centerline{\includegraphics[width=8.4cm]{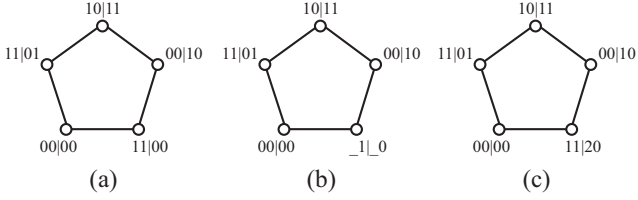}}
%\begin{figure}[t] \centerline{\includegraphics[width=8.4cm]{Figures\Fig1.pdf}}
\caption{\label{pentagonfirst} Events corresponding to the first (a), second (b), and third (c) pentagonal Bell inequalities.}
\end{figure}

%%%%%%%%%%%%%%%%%%%%%%%%%%%%%%%%%%%%%%%%%%%%%%%%%%%%%%%%%%%%%%%%%%%

By inspection (see Appendix~\ref{App1}), we found that there are three nonequivalent Bell inequalities associated with a pentagon.

%%%%%%%%%%%%%%%%%%%%%%%%%%%%%%%%%%%%%%%%%%%%%%%%%%%%%%%%%%%%%%%%%%%

\section{First Bell inequality}

%%%%%%%%%%%%%%%%%%%%%%%%%%%%%%%%%%%%%%%%%%%%%%%%%%%%%%%%%%%%%%%%%%%

One set of local events satisfying the constraints of the pentagon is shown in Fig.~\ref{pentagonfirst}(a). The corresponding Bell inequality
\begin{equation}
\begin{split}
 &P(00|00)+P(11|01)+P(10|11)+P(00|10)\\
 &+P(11|00) \stackrel{\mbox{\tiny{LHV}}}{\leq} 2 \stackrel{\mbox{\tiny{QM}}}{\leq} 2.178
 \label{first}
 \end{split}
\end{equation}
is a two-setting, two-outcome (both for Alice and Bob) Bell inequality. Its local bound is $\Omega_\mathrm{LHV}=\alpha(C_5)=2$, and its quantum bound $\Omega_\mathrm{QM} \approx 2.178$ (see Appendix~\ref{App3}) occurs for states locally equivalent to
$
|\psi\rangle = 0.6338 |00\rangle + 0.7735 |11\rangle.
$
The resulting optimal probabilities are
$P(00|00) = |(c,-s) \otimes (-s,c)|\psi\rangle|^2$,
$P(11|01) = |(s,c) \otimes (-C,-S) |\psi\rangle|^2$,
$P(10|11) = |(S,C) \otimes (-S,C) |\psi\rangle|^2$,
$P(00|10) = |(C,-S) \otimes (-s,c) |\psi\rangle|^2$, and
$P(11|00) = |(s,c) \otimes (c,s) |\psi\rangle|^2$,
respectively, with $c=0.7911$, $s=0.6117$, $C=0.2152$, and $S=0.9766$.
Notice that the inequality is maximally violated by nonmaximally entangled sates.

%%%%%%%%%%%%%%%%%%%%%%%%%%%%%%%%%%%%%%%%%%%%%%%%%%%%%%%%%%%%%%%%%%%

\section{Second Bell inequality}

%%%%%%%%%%%%%%%%%%%%%%%%%%%%%%%%%%%%%%%%%%%%%%%%%%%%%%%%%%%%%%%%%%%

Relaxation of the symmetry between Alice's and Bob's actions in \eqref{first} allows one to substitute Alice's measurement in
vertex $11|00$ of Fig.~\ref{pentagonfirst}(a) with the identity. The substitution produces a set of events depicted in Fig.~\ref{pentagonfirst}(b) and a new Bell inequality
\begin{equation}
\begin{split}
 &P(00|00)+P(11|01)+P(10|11)+P(00|10)\\
 &+P(\underline{\;\;}1|\underline{\;\;}0) \stackrel{\mbox{\tiny{LHV}}}{\leq} 2 \stackrel{\mbox{\tiny{QM}}}{\leq} \frac{3+\sqrt{2}}{2} \approx 2.207,
 \label{second}
\end{split}
\end{equation}
where $P(\underline{\;\;}1|\underline{\;\;}0)$ denotes the probability of Bob obtaining 1 when measuring in setting 0 irrespectively of Alice's action.

In this case, the maximum quantum value is achieved with maximally
entangled states, e.g., with
$|\phi^+\rangle =
\frac{1}{\sqrt{2}} (|00\rangle + |11\rangle)$.
The optimal probabilities are
$P(00|00) = |(0,1) \otimes (-s,c) |\phi^+\rangle|^2$,
$P(11|01) = |(1,0) \otimes (-c,s) |\phi^+\rangle|^2$,
$P(10|11) = |(1/\sqrt{2},1/\sqrt{2}) \otimes (s,c) |\phi^+\rangle|^2$, and
$P(00|10) = |(-1/\sqrt{2},1/\sqrt{2}) \otimes (-s,c) |\phi^+\rangle|^2$,
with $c=\cos\left(\frac{\pi}{8}\right)$ and $s=\sin\left(\frac{\pi}{8}\right)$. In an ideal experiment, the first four probabilities are $\frac{c^2}{2} \approx 0.427$, while $P(\underline{\;\;}1|\underline{\;\;}0)=\frac{1}{2}$.

%%%%%%%%%%%%%%%%%%%%%%%%%%%%%%%%%%%%%%%%%%%%%%%%%%%%%%%%%%%%%%%%%%%

\section{Third Bell inequality}

%%%%%%%%%%%%%%%%%%%%%%%%%%%%%%%%%%%%%%%%%%%%%%%%%%%%%%%%%%%%%%%%%%%

Without changing the maximum quantum violation, one can substitute the optimal one-dimensional projection for the identity measurement in Fig.~\ref{pentagonfirst}(b). This leads to Fig.~\ref{pentagonfirst}(c) and to the following three-setting (for Alice), two-setting (for Bob), two-outcome Bell inequality:
\begin{equation}
\begin{split}
 &P(00|00)+P(11|01)+P(10|11)+P(00|10)\\
 &+P(11|20) \stackrel{\mbox{\tiny{LHV}}}{\leq} 2 \stackrel{\mbox{\tiny{QM}}}{\leq} \frac{3+\sqrt{2}}{2}.
 \label{third}
\end{split}
\end{equation}
The optimal local tests leading to the maximum quantum violation are the same as those of inequality \eqref{second}, with Alice's additional measurement
projecting on $(c,s)$ and producing $P(11|20)=|(c,s) \otimes (c,s) |\phi^+\rangle|^2$.

Inequalities \eqref{first}, \eqref{second}, and \eqref{third} exhaust the list of the Bell inequalities extractable from a pentagon. All three inequalities are maximally violated on pairs of qubits (see Appendix~\ref{App2}).

%%%%%%%%%%%%%%%%%%%%%%%%%%%%%%%%%%%%%%%%%%%%%%%%%%%%%%%%%%%%%%%%%%%
% Fig. 2
%%%%%%%%%%%%%%%%%%%%%%%%%%%%%%%%%%%%%%%%%%%%%%%%%%%%%%%%%%%%%%%%%%%

\begin{figure}[t] \centerline{\includegraphics[width=4.8cm]{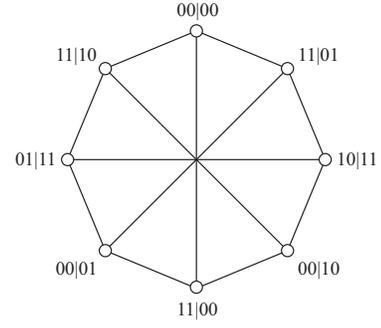}}
%\begin{figure}[t] \centerline{\includegraphics[width=5.2cm]{Figures\Fig2.pdf}}
\caption{\label{CHSH}Graph of the relationships of exclusivity between the eight events in the inequality \eqref{CHSH3}.}
\end{figure}

%%%%%%%%%%%%%%%%%%%%%%%%%%%%%%%%%%%%%%%%%%%%%%%%%%%%%%%%%%%%%%%%%%%

\section{Connection to other Bell inequalities}

%%%%%%%%%%%%%%%%%%%%%%%%%%%%%%%%%%%%%%%%%%%%%%%%%%%%%%%%%%%%%%%%%%%

Probabilities $P(ab|xy)$ can be related to the expectation values of dichotomic observables $A_x$ and $B_y$ with possible results $-1$ and $+1$, as $P(ab|xy) = \frac{1}{4} \langle [\openone+(-1)^{a}A_x][\openone+(-1)^{b}B_y]\rangle$. This allows one to rewrite inequality \eqref{second} as $\frac{3}{2} + \frac{1}{4} (\langle A_0B_0 \rangle + \langle A_0B_1 \rangle + \langle A_1B_0 \rangle - \langle A_1B_1 \rangle )\stackrel{\mbox{\tiny{LHV}}}{\leq} 2 \stackrel{\mbox{\tiny{QM}}}{\leq} \frac{3+\sqrt{2}}{2}$, which is a rescaled and shifted CHSH inequality $\langle A_0B_0 \rangle + \langle A_0B_1 \rangle + \langle A_1B_0 \rangle - \langle A_1B_1 \rangle \stackrel{\mbox{\tiny{LHV}}}{\leq} 2 \stackrel{\mbox{\tiny{QM}}}{\leq} 2\sqrt{2}$.

A direct link between inequality \eqref{second} and CHSH provides a particularly simple argument for the impossibility of PR nonlocal boxes \cite{PR94}. It has been recently shown \cite{A12,Cabello12} that the maximum bound in a pentagonal inequality satisfying the condition that the sum of the probabilities of exclusive events does not exceed $1$ is $\sqrt{5}$. When applied to inequality \eqref{second}, this translates into $\langle A_0B_0 \rangle + \langle A_0B_1 \rangle + \langle A_1B_0 \rangle - \langle A_1B_1 \rangle \leq 4 \sqrt{5} - 6 \approx 2.944$, which is considerably lower than $4$, which is the value obtained with PR boxes.

The other two pentagonal Bell inequalities are not tight inequalities. Nevertheless, inequalities \eqref{first} and \eqref{third} are directly related to CHSH and $I_{3322}$, respectively.

A sum of events in eight inequalities equivalent to \eqref{first} is represented by the $(1,4)$-circulant graph on eight vertices, $Ci_8(1,4)$, shown in Fig.~\ref{CHSH}. The resulting inequality reads
\begin{equation}
\label{CHSH3}
 \sum P(a,b|x,y) \stackrel{\mbox{\tiny{LHV}}}{\leq} 3 \stackrel{\mbox{\tiny{QM}}}{\leq} 2+\sqrt{2},
\end{equation}
where the sum is extended to all $x,y\in \{0,1\}$ and $a,b\in \{0,1\}$ such that $a \oplus b=xy$, where $\oplus$ denotes sum modulo $2$. Like for inequality \eqref{second}, it is a simple exercise to show that \eqref{CHSH3} is nothing else but the CHSH inequality written in terms of a convex combination of probabilities. Thus inequality \eqref{first} can be regarded as a primitive building block in the CHSH inequality. An interesting fact is that, while the quantum maximum for inequality \eqref{second} is strictly below the Lov\'asz number of $C_5$, the quantum bound for \eqref{CHSH3} is equal to the Lov\'asz number of $Ci_8(1,4)$.

Finally, inequality \eqref{third} is induced in the $I_{3322}$ inequality, a Bell inequality with three dichotomic tests for Alice and for Bob, which can be written as
\begin{equation}
\begin{split}
&P(11|00)+P(11|01)+P(00|10)+P(10|11) \\ &+P(00|02)+P(00|20)+P(00|21)+P(10|22) \\
&+P(\underline{\;\;}1|\underline{\;\;}2) + P(1\underline{\;\;}|2\underline{\;\;})
\stackrel{\mbox{\tiny{LHV}}}{\leq} 4.
\end{split}
\end{equation}
This expression contains the five terms of the pentagonal Bell inequality~\eqref{third}, which again illustrates that $I_{3322}$ contains logically simpler pentagonal inequalities.

%%%%%%%%%%%%%%%%%%%%%%%%%%%%%%%%%%%%%%%%%%%%%%%%%%%%%%%%%%%%%%%%%%%

\section{Pentagonal Bell inequalities vs pentagonal NC inequalities}

%%%%%%%%%%%%%%%%%%%%%%%%%%%%%%%%%%%%%%%%%%%%%%%%%%%%%%%%%%%%%%%%%%%

Bell inequalities can be regarded as noncontextualilty inequalities in which compatible tests are performed on spatially separated subsystems. Specifically, this restriction implies that in bipartite Bell experiments there are three possible types of relationships of exclusivity between two events:
(i) exclusivity between Alice's events only (e.g., between $00|00$ and $11|01$); (ii) exclusivity between Bob's events only (e.g., between $00|00$ and $11|10$); and (iii) exclusivity between Alice's and Bob's events (e.g., between $00|00$ and $11|00$).

To encode this information, the edges of a graph representing a bipartite Bell inequality must be of three different types. However, the Lov\'asz number of a graph is insensitive to this extra information.
It is thus not surprising that the Lov\'asz number of the pentagon, $\vartheta(C_5)$, gives only a loose upper bound for the violations of the pentagonal Bell inequalities \eqref{first}, \eqref{second}, and \eqref{third}, while for the KCBS inequality \cite{KCBS08},
\begin{equation}
 \begin{split}
 &P(01|01)+P(01|12)+P(01|23)+P(01|34)\\
 &+P(01|40) \stackrel{\mbox{\tiny{NCHV}}}{\leq} 2 \stackrel{\mbox{\tiny{QM}}}{\leq} \vartheta(C_5)=\sqrt{5} \approx 2.236.
 \label{pK}
 \end{split}
\end{equation}
The important point is that while $\sqrt{5}$ is physically accessible \cite{SM2}, that is not the case if we replace compatibility by spacelike separation. This shows that, from the point of view of QM, the extra requirement of spacelike separation adds mathematical complexity to a problem which, without this requirement, has a simple solution, since the Lov\'asz number of the graph can be calculated with a semidefinite program.

It is an open question whether by introducing different types of edges corresponding to relationships of exclusivity of types (i), (ii), and (iii), respectively, one can utilize graph theory to design efficient algorithms for calculations of the tight bounds for Bell inequalities.
This approach may be an alternative to the one based on solving the hierarchy of semidefinite programs \cite{NPA08}.

%%%%%%%%%%%%%%%%%%%%%%%%%%%%%%%%%%%%%%%%%%%%%%%%%%%%%%%%%%%%%%%%%%%

\section{Experimental tests}

%%%%%%%%%%%%%%%%%%%%%%%%%%%%%%%%%%%%%%%%%%%%%%%%%%%%%%%%%%%%%%%%%%%

To test the Bell inequalities experimentally, we used pairs of photons entangled in polarization. The main objectives were to demonstrate that quantum nonlocality can be observed directly via the elementary (pentagonal) inequalities and to provide tests of pentagonal NC inequalities free of the compatibility loophole \cite{GKCLKZGR10}.

%%%%%%%%%%%%%%%%%%%%%%%%%%%%%%%%%%%%%%%%%%%%%%%%%%%%%%%%%%%%%%%%%%%
% Fig. 3
%%%%%%%%%%%%%%%%%%%%%%%%%%%%%%%%%%%%%%%%%%%%%%%%%%%%%%%%%%%%%%%%%%%

\begin{figure}[t] %\vspace{3.4cm}
 \centerline{\includegraphics[width=7.2cm]{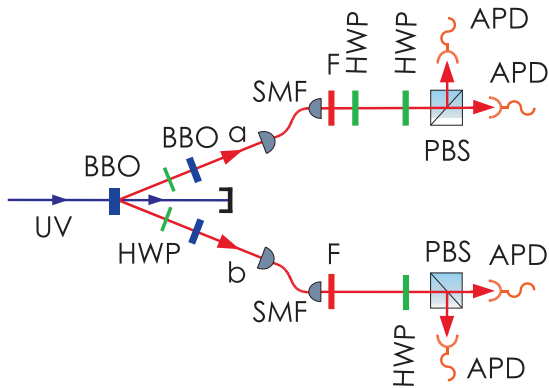}}
 %\centerline{\includegraphics[width=7.0cm]{Figures\Fig3.pdf}}
 \caption{\label{Setup} Experimental setup for testing the three pentagonal Bell inequalities.}
\end{figure}

%%%%%%%%%%%%%%%%%%%%%%%%%%%%%%%%%%%%%%%%%%%%%%%%%%%%%%%%%%%%%%%%%%%

The experiment is shown in Fig.~\ref{Setup}: Ultraviolet light centered at a wavelength of 390 nm were focused inside a 2-mm-thick $\beta$ barium borate (BBO) nonlinear crystal, to produce photon pairs emitted into two spatial modes $a$ and $b$ through the second order degenerate emission of type-II spontaneous parametric down-conversion. Half wave plates (HWP) and two 1-mm-thick BBO crystals were used for compensation of longitudinal and transversal walk-offs. The emitted photons were coupled into 2 m single-mode optical fibers (SMF) and passed through a narrow-bandwidth interference filters ($F$) ($\Delta\lambda=1$ nm) to secure well-defined spatial and spectral emission modes. To observe $|\phi_{+}\rangle$ and $|\psi\rangle$, a HWP was placed after the output fiber coupler in mode ($a$). The polarization measurement was performed using HWPs and polarizing beam splitters (PBS) followed by actively quenched Si-avalanche photodiodes (Si-APD).

The measurement time for each pair of local tests was $100$ s. The results for the Bell inequalities \eqref{first}, \eqref{second}, and \eqref{third} are presented in Tables \ref{TableI}, \ref{TableII}, and \ref{TableIII}, respectively. The column labeled ``Ideal'' in Tables \ref{TableI}, \ref{TableII}, and \ref{TableIII} contains the theoretical predictions for the ideal case in which the prepared state and the local tests are exactly those needed to obtain the maximum in QM.

%%%%%%%%%%%%%%%%%%%%%%%%%%%%%%%%%%%%%%%%%%%%%%%%%%%%%%%%%%%%%%%%%%%
% Table I
%%%%%%%%%%%%%%%%%%%%%%%%%%%%%%%%%%%%%%%%%%%%%%%%%%%%%%%%%%%%%%%%%%%

\begin{table}[tb]
 \caption{\label{TableI}Experimental results for the test of inequality \eqref{first}.}
\begin{ruledtabular}
{ \begin{tabular}{l c l}
 Correlation & Experimental & Ideal\\
\hline
 $P(00|00)$ & $0.4407 \pm 0.0072$ & $0.464$ \\
 $P(11|01)$ & $0.4718 \pm 0.0076$ & $0.464$ \\
 $P(10|11)$ & $0.3117 \pm 0.0075$ & $0.323$ \\
 $P(00|10)$ & $0.4719 \pm 0.0078$ & $0.464$ \\
 $P(11|00)$ & $0.4423 \pm 0.0081$ & $0.464$ \\
 $\Omega_{\rm QM}$ & $2.138 \pm 0.017$ & $2.178$ \\
 \end{tabular} }
\end{ruledtabular}
\end{table}

%%%%%%%%%%%%%%%%%%%%%%%%%%%%%%%%%%%%%%%%%%%%%%%%%%%%%%%%%%%%%%%%%%%
% Table II
%%%%%%%%%%%%%%%%%%%%%%%%%%%%%%%%%%%%%%%%%%%%%%%%%%%%%%%%%%%%%%%%%%%

\begin{table}[tb]
 \caption{\label{TableII}Experimental results for the test of inequality \eqref{second}.}
 \begin{ruledtabular}
 { \begin{tabular}{l c l}
 Correlation & Experimental & Ideal\\
\hline
 $P(00|00)$ & $0.4215 \pm 0.0075$ & $0.427$ \\
 $P(11|01)$ & $0.4313 \pm 0.0067$ & $0.427$ \\
 $P(10|11)$ & $0.4307 \pm 0.0077$ & $0.427$ \\
 $P(00|10)$ & $0.4228 \pm 0.0070$ & $0.427$ \\
 $P(\underline{\;\;}1|\underline{\;\;}0)$ & $0.5019 \pm 0.0059$ & $0.5$ \\
 $\Omega_{\rm QM}$ & $2.208 \pm 0.016$ & $2.207$ \\
\end{tabular} }
\end{ruledtabular}
\end{table}

%%%%%%%%%%%%%%%%%%%%%%%%%%%%%%%%%%%%%%%%%%%%%%%%%%%%%%%%%%%%%%%%%%%
% Table III
%%%%%%%%%%%%%%%%%%%%%%%%%%%%%%%%%%%%%%%%%%%%%%%%%%%%%%%%%%%%%%%%%%%

\begin{table}[tb]
 \caption{\label{TableIII}Experimental results for the test of inequality \eqref{third}.}
\begin{ruledtabular}
{ \begin{tabular}{l c l}
 Correlation & Experimental & Ideal\\
 \hline
 $P(00|00)$ & $0.4215 \pm 0.0075$ & $0.427$ \\
 $P(11|01)$ & $0.4313 \pm 0.0067$ & $0.427$ \\
 $P(10|11)$ & $0.4307 \pm 0.0077$ & $0.427$ \\
 $P(00|10)$ & $0.4228 \pm 0.0070$ & $0.427$ \\
 $P(11|20)$ & $0.4999 \pm 0.0079$ & $0.5$ \\
 $\Omega_{\rm QM}$ & $2.206 \pm 0.017$ & $2.207$ \\
\end{tabular} }
\end{ruledtabular}
\end{table}

%%%%%%%%%%%%%%%%%%%%%%%%%%%%%%%%%%%%%%%%%%%%%%%%%%%%%%%%%%%%%%%%%%%

\section{Conclusions}

%%%%%%%%%%%%%%%%%%%%%%%%%%%%%%%%%%%%%%%%%%%%%%%%%%%%%%%%%%%%%%%%%%%

We identified and experimentally tested the three bipartite Bell inequalities with relationships of exclusivity between events given by a pentagon, the simplest graph with a classical-quantum gap.
Our inequality \eqref{second} is algebraically equivalent to CHSH inequality; the pentagonal structures of inequalities \eqref{first} and \eqref{third} appear as building blocks in a more traditional form of CHSH and in $I_{3322}$, respectively. Thus, the quantum violation of the latter inequalities could be traced back to the quantum violation in their elementary logical components. Moreover, by directly linking inequality \eqref{second} to CHSH we could provide a particularly simple argument for the impossibility of PR nonlocal boxes.

The maximum quantum violation of a given graph of the relationships of exclusivity is easily computable. The same task for a Bell inequality is, on the other hand, generally hard. We linked this apparent paradox to the fact that the constraints imposed by a bipartite scenario introduce three possible types of relationships of exclusivity, which in turn would require a graph with three different types of edges, thus adding computational complexity to an otherwise simple problem.

The experimental tests of the three inequalities show that quantum nonlocality can actually be observed through the violation of any of these elementary logical Bell inequalities. In addition, our experiments show how pentagonal NC inequalities can be violated without the compatibility loophole \cite{GKCLKZGR10} affecting previous experiments carried out by performing sequential tests on the same system \cite{LLSLRWZ11, AABC11}. In our experiments, the spatial separation between the first and the second test guarantees perfect compatibility without further assumptions. In this sense, our experiments can be considered tests of pentagonal NC inequalities free of the compatibility loophole.

%%%%%%%%%%%%%%%%%%%%%%%%%%%%%%%%%%%%%%%%%%%%%%%%%%%%%%%%%%%%%%%%%%%

\begin{acknowledgments}
The authors thank J.-\AA. Larsson, M. Navascu\'es, S.~Severini, T. V\'{e}rtesi, and A. Winter for stimulating discussions. This work was supported by the Pakistani Higher Education Commission, the Swedish Research Council (VR), the Spanish Projects No.\ FIS2008-05596 and No.\ FIS2011-29400, and the Wenner-Gren Foundation.
\end{acknowledgments}

%%%%%%%%%%%%%%%%%%%%%%%%%%%%%%%%%%%%%%%%%%%%%%%%%%%%%%%%%%%%%%%%%%%

\appendix

\section{There are only three bipartite pentagonal Bell inequalities}
\label{App1}

%%%%%%%%%%%%%%%%%%%%%%%%%%%%%%%%%%%%%%%%%%%%%%%%%%%%%%%%%%%%%%%%%%%

In a bipartite scenario, the relationships of exclusive disjunction between events (represented by edges linking vertices) can be due to two reasons: because Alice's local events are exclusive (in this case, we will call the corresponding edge an $A$ edge) or because Bob's local events are exclusive ($B$ edge). For simplicity, we regard the case in which both Alice's and Bob's local events are exclusive as either $A$ or $B$. Correspondingly, vertices are of three types: $AA$, when the vertex has an $A$ edge with one neighbor and an $A$ edge with the other neighbor, $AB$, or $BB$. With this notation, there are four essentially different edge patterns in a pentagon: those containing four $AB$ vertices (edge pattern $BABAB$), those containing two $AB$ vertices (edge patterns $BBBAA$ and $BBBBA$), and those containing five $BB$ vertices (edge pattern $BBBBB$).

It is easy to see that any edge pattern containing $BBB$ is not compatible with a pentagonal bipartite Bell inequality: If vertex $v_0$ corresponds to Bob's $0|0$ and vertex $v_1$ corresponds to Bob's $1|0$ then, in order not to be exclusive with vertex $v_0$, vertex $v_2$ may not be Bob's $0|0$ (it may be, e.g., $0\;{\rm or}\; 2|0$). This implies that vertex $3$ must exclude Bob's $0|0$. Vertices $0$ and $3$ are thus exclusive. However, this is not allowed if the structure is a pentagon. Consequently, $BBABA$ is the only edge pattern which may represent a pentagonal Bell inequality.

\begin{figure}[t]
 \centerline{\includegraphics[width=4.2cm]{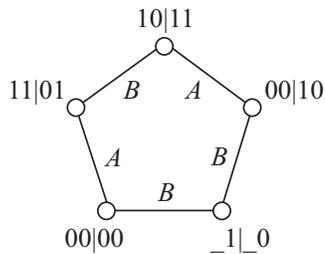}}
%\begin{figure}[t] \centerline{\includegraphics[width=8.4cm]{Figures\Fig1.pdf}}
\caption{\label{FigSM1} Assignments fixed by an edge pattern $BBABA$.}
\end{figure}

This pattern fixes the tests and results to those depicted in Fig.~\ref{FigSM1}. There, Alice's role in vertex $BB$ is not defined. Therefore, she may either do nothing, or perform a nontrivial measurement provided that its result does not create any additional edge. Alice doing nothing leads to Bell inequality \eqref{second}, and Alice performing each of two possible nontrivial measurements leads to Bell inequalities \eqref{first} and \eqref{third}.

%%%%%%%%%%%%%%%%%%%%%%%%%%%%%%%%%%%%%%%%%%%%%%%%%%%%%%%%%%%%%%%%%%%
%Fig. 1
%%%%%%%%%%%%%%%%%%%%%%%%%%%%%%%%%%%%%%%%%%%%%%%%%%%%%%%%%%%%%%%%%%%

%%%%%%%%%%%%%%%%%%%%%%%%%%%%%%%%%%%%%%%%%%%%%%%%%%%%%%%%%%%%%%%%%%

\section{Two qubits are enough to reach the maximum quantum violation}
\label{App2}

%%%%%%%%%%%%%%%%%%%%%%%%%%%%%%%%%%%%%%%%%%%%%%%%%%%%%%%%%%%%%%%%%%

In an inequality, like our inequalities \eqref{first} and \eqref{second}, containing not more than two projectors for Alice, the Bell operator can be written as follows:
\begin{equation}
 S = P_1 \otimes Q_1 + P_2 \otimes Q_2 + \openone \otimes Q_0,
\end{equation}
where $P_1$ and $P_2$ are projections on subspaces of Alice, $\openone$ is Alice's identity, and $Q_1$ and $Q_2$ are, in principle, arbitrary observables at Bob's side. Thus one can write
\begin{equation}
 P_1 =\sum_i |e_i\rangle\langle e_i| \ \ \textrm{and} \ \ P_2 =\sum_j |f_j\rangle\langle f_j|,
\end{equation}
where ${|e_i\rangle}$ and ${|f_j\rangle}$ are (for now arbitrary) orthonormal bases in the span of $P_1$ and $P_2$, respectively. Scalar products of the vectors in the two bases form matrix $G$ with elements $G_{ij} = \langle e_i|f_j\rangle$.

Now, let the two unitaries $U$ and $V$ produce a singular value decomposition of $G$, i.e., let
\begin{equation}
 D_{ij} =\sum_{km} U^{\dag}_{ik}G_{km}V_{mj} = \lambda_i \delta_{ij}.
\end{equation}
The unitaries rotate the base vectors into $|p_{\mu}^1\rangle = \sum_j |e_j\rangle U_{j\mu}$ and $|p_{\mu}^2\rangle = \sum_j |f_j\rangle V_{j\mu}$, with the property $\langle p_{\mu}^1|p_{\nu}^2\rangle = \lambda_{\mu} \delta_{\mu \nu}$.

Due to the orthogonality of $|p_{\mu}^1\rangle$ and $|p_{\nu}^2\rangle$ for $\mu \neq \nu$, the Bell operator splits into the following blocks along its main diagonal:
\begin{equation}
\label{BlockA}
 S_{\mu} =|p_{\mu}^1\rangle\langle p_{\mu}^1| \otimes Q_1 + |p_{\mu}^2\rangle\langle p_{\mu}^2| \otimes Q_2 + \openone_{\mu} \otimes Q_0,
\end{equation}
where $\openone_{\mu}$ denotes the identity operator in a maximally two-dimensional space spanned by $|p_{\mu}^1\rangle$ and $|p_{\mu}^2\rangle$.

Each block \eqref{BlockA} can now be rewritten in a form where the projections of $Q_1$ and $Q_2$ are factored out. Then, one can repeat the procedure exchanging the roles of Alice and Bob. Then, the procedure produces a block-diagonal form of the Bell operator with no more than two-dimensional subspaces of Alice and of Bob in a block. Optimization of one such block yields the maximum eigenvalue of the Bell operator.

%%%%%%%%%%%%%%%%%%%%%%%%%%%%%%%%%%%%%%%%%%%%%%%%%%%%%%%%%%%%%%%%%%%

\section{Optimal quantum violation of the pentagonal inequalities}
\label{App3}

%%%%%%%%%%%%%%%%%%%%%%%%%%%%%%%%%%%%%%%%%%%%%%%%%%%%%%%%%%%%%%%%%%%

To optimize the quantum violation of the Bell inequality \eqref{first} we have taken into account that two qubits are enough to reach the quantum maximum, assuming that $A_0$ and $B_0$ are the Pauli matrix $\sigma_z$, and that $A_1$ and $B_1$ are linear combinations of $\sigma_z$ and $\sigma_x$. Then, the resulting Bell operator is a function of two real parameters. The maximum violation of the inequality is simply given by the largest eigenvalue of the Bell operator. The eigenvalue problem leads to a nontrivial equation of degree three whose maximal root has to be maximized. The maximization does not lead to any result in a concise analytic form. For this reason, in this case we have presented numerical results.

The maximum quantum violation of inequalities \eqref{second} and \eqref{third} easily follows from the connection between inequality \eqref{second} and the CHSH inequality, and from the connection between inequalities \eqref{second} and \eqref{third}.

%%%%%%%%%%%%%%%%%%%%%%%%%%%%%%%%%%%%%%%%%%%%%%%%%%%%%%%%%%%%%%%%%%%


\begin{thebibliography}{99}
\bibitem{SM2}
 The maximum quantum value can be obtained preparing the state
%\begin{equation}
$\langle\psi|=(0,0,1)$
%\end{equation}
and with the tests $i=2
|v_i\rangle \langle v_i|-\openone$ with
%\begin{subequations}
%\label{vectors}
%\begin{align}
$\langle v_{0}| = N_{0}\left(1,0,r\right)$,
$\langle v_{1,4}| = N_{1}\left(c,\pm s,r\right)$,
$\langle v_{2,3}| = N_{2}\left(C,\mp S,r\right)$,
%\end{align}
%\end{subequations}
where
$r=\sqrt{\cos{\left(\frac{\pi}{5}\right)}}$,
$c=\cos{\left(\frac{4\pi}{5}\right)}$, $s=\sin{\left(\frac{4\pi}{5}\right)}$,
$C=\cos{\left(\frac{2\pi}{5}\right)}$, $S=\sin{\left(\frac{2\pi}{5}\right)}$,
and $N_{i}$ are suitable normalization factors.
\end{thebibliography}
\end{document}